\documentclass[aps,prb,reprint,superscriptaddress,longbibliography,floatfix]{revtex4-2}
\usepackage[utf8]{inputenc}
\usepackage[T1]{fontenc}
\usepackage{amsmath,amssymb,bm,mathtools}
\usepackage{graphicx,booktabs}
\usepackage{xcolor}
\usepackage{array}
\newif\ifshowrevisions
\showrevisionstrue
\newcommand{\revstart}{}
\newcommand{\revstop}{}

\makeatletter
\@ifpackageloaded{ulem}{}{\IfFileExists{ulem.sty}{\usepackage[normalem]{ulem}}{}}
\makeatother

\ifdefined\sout
  
\else
  
\fi

\usepackage[colorlinks=true,allcolors=blue]{hyperref}
\DeclareMathOperator{\sgn}{sgn}

\newcommand{\SM}{Supplemental Material}
\hypersetup{pdftitle={Maximal Chern Numbers from Finite-Range Hopping}}

\begin{document}
\title{Maximal Chern Numbers from Finite-Range Hopping}
\author{Christopher Arriagada Cortés}
\affiliation{Departamento de F\'isica, Universidad T\'ecnica Federico Santa Mar\'ia, Casilla 110, Valpara\'iso, Chile}
\author{Vladimir Juri\v ci\'c}\thanks{Corresponding author:vladimir.juricic@usm.cl}
\affiliation{Departamento de F\'isica, Universidad T\'ecnica Federico Santa Mar\'ia, Casilla 110, Valpara\'iso, Chile}
\begin{abstract}
Finite-range hopping constrains both the number of Dirac points and the masses that gap them. We determine the maximal absolute Chern numbers of an isotropic finite-range extension of the Qi--Wu--Zhang model on a square lattice through thirteenth-neighbor hopping. Algebraic zero counting and mass-weighted sum rules yield global bounds, and we construct explicit Hamiltonians that saturate them. We find $|C|_{\max}=36,40,51$ for eleventh-, twelfth-, and thirteenth-neighbor hopping. Equal numbers of Dirac points can produce different maxima because finite-range mass harmonics distinguish opposite vortex charges differently. Twice the squared hopping radius bounds every maximum through twelfth-neighbor hopping and is saturated at the eleventh and twelfth, but the thirteenth-neighbor maximum exceeds it. Thus finite-range hopping constrains the maximal Chern number in two
complementary ways: through the number of Dirac vortices it can generate
and through the mass-sign patterns it can assign to them. This geometric
viewpoint provides a design principle for high-Chern bands and a route to
analogous bounds in moir\'e and other lattice systems.
\end{abstract}
\maketitle

\textit{Introduction.---}
The Chern number determines the quantized Hall response of a
two-dimensional band insulator~\cite{TKNN1982,Berry1984} and the net
number of chiral edge modes~\cite{Hatsugai1993}. Its realization
without net magnetic flux~\cite{Haldane1988} established Chern bands
as a central platform for topological quantum
matter~\cite{HasanKane2010,QiZhang2011}. Nearly flat Chern
bands~\cite{Tang2011,Sun2011} provide a lattice setting for
fractional quantum Hall physics~\cite{Kapit2010,Neupert2011,RegnaultBernevig2011}, while
bands with $|C|>1$ support multiple propagating edge channels and
correlated phases with richer internal structure~
\cite{Wang2012,Liu2012,Moller2015}. Higher and even arbitrary Chern
numbers can be engineered in lattice models~
\cite{Yang2012,Fan2026}, with high-Chern bands realized in settings
including $\beta$-graphyne, dice and star lattices, pyrochlore slabs,
and kagome ferromagnets~
\cite{vanMiert2014,WangRan2011,Chen2012,Trescher2012,Zhang2021}.
These developments raise a basic question: how large can the Chern
number become when the orbital content and hopping range are fixed?

Moir\'e materials motivate this question because orbital structure,
interference, and symmetry can make hopping processes at different
distances compete~\cite{Koshino2018,Kang2018,Crepel2024,Eugenio2025},
while twisted multilayers can host flat bands with higher Chern
numbers~\cite{Ledwith2022}.
Their narrow bands host correlated
phases~\cite{Bistritzer2011,Cao2018a,Cao2018b}, while twist and
electric fields tune bandwidths and
topology~\cite{Zhang2019,Wu2018,Wu2019,Devakul2021}.
Experiments reveal anomalous~\cite{Sharpe2019} and
quantized~\cite{Serlin2020,Chen2020} Hall responses, as well as
fractional Chern states at finite field in
graphene~\cite{Xie2021} and at zero field in twisted
MoTe$_2$~\cite{Cai2023,Zeng2023,Park2023}.
The competing hopping channels motivate determining what their
spatial support permits before imposing material-specific
relations among the amplitudes.

Models with arbitrary Chern number have been constructed through
orbital and momentum-space approaches~\cite{Yang2012,Trescher2012,Fan2026}.
In two-band systems, the signed-vortex description relates Chern number to
vortex charges of Dirac points and the corresponding masses~\cite{Sticlet2012};
further-neighbor honeycomb models reach $|C|=5$, with larger
values appearing to require longer-range mass terms~\cite{Sticlet2013}.
Previous work has connected hopping range to the attainable Chern
number~\cite{Udagawa2014} and to the optimization of band flatness and
geometry under finite-range hopping~\cite{Lee2016,LeeClaassenThomale2017}.
Related locality constraints have been developed for Wannier
representations and exactly flat topological bands~
\cite{Marzari2012,Brouder2007,DubailRead2015,Chen2014,Read2017}.

\revstart
\begin{figure*}[t]
\centering
\includegraphics[width=\textwidth]{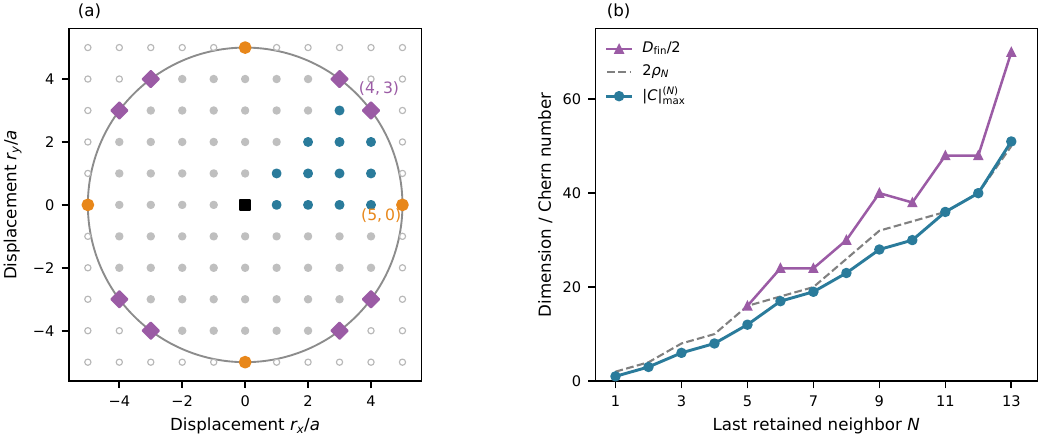}
\caption{Hopping geometry and maximal Chern numbers.
(a) Sites within the cutoff $R=5a$ (circle) around the origin
(black square); open symbols lie outside. Blue marks the twelve inner orbit representatives,
filled gray their partners, and orange and purple the independent
$(5,0)$ and $(4,3)$ orbits at $\rho_{13}=25$.
(b) Maximal $|C|$ for Eqs.~\eqref{eq:model}--\eqref{eq:harmonics},
compared with $D_{\rm fin}/2$ for $N=5,\ldots,13$ and $2\rho_N$.
Here $D_{\rm fin}$ counts finite complex hybridization zeros.
Lines connecting symbols guide the eye. Proofs and saturating
Hamiltonians are given in Secs.~SIII--SXVI of
SM~\cite{SupplementalMaterial}.}
\label{fig:support}
\end{figure*}

We determine the maximal Chern numbers of the isotropic two-orbital
square-lattice family of models through thirteenth-neighbor hopping. At fixed
hybridization, the available mass harmonics control how the Dirac
vortices can be gapped, while mass-weighted sum rules impose bounds
that hold throughout the full gapped parameter space. These bounds are
sharp at every hopping range considered, with
$|C|_{\max}^{(11)}=36$, $|C|_{\max}^{(12)}=40$, and
$|C|_{\max}^{(13)}=51$. Twice the squared hopping radius bounds all
maxima through twelfth-neighbor hopping and is saturated at the
eleventh and twelfth, whereas the thirteenth-neighbor maximum exceeds
this envelope. The \SM{} (SM)~\cite{SupplementalMaterial} contains
the proofs and explicit saturating Hamiltonians.

\revstart
\textit{Lattice model.---}
We consider a finite-range extension of the Qi--Wu--Zhang model (one Chern block of the Bernevig--Hughes--Zhang Hamiltonian) on a square lattice~\cite{QWZ-PRB-2006,Bernevig2006},
% APS display rows: 1
\begin{equation}
 H_N(\mathbf k)=X_N\sigma_x-Y_N\sigma_y+m_N\sigma_z,
 \label{eq:model}
\end{equation}
where the Pauli matrices act on the two orbitals. The complex
amplitude $g_N\equiv X_N+iY_N$ hybridizes them, while $m_N$ describes their relative
onsite energy and diagonal hopping. The resulting band topology is
characterized by an integer Chern invariant~\cite{QWZ-PRB-2006}.
We set the lattice constant to unity and retain the displacement
representatives
$\mathcal D_N=\{(a,b):a\ge b\ge0,\ 0<a^2+b^2\le\rho_N\}$,
with $\rho_N$ the squared radius of the $N$th neighbor shell.
The momentum dependence is
% APS display rows: 3
\begin{align}
 X_N(k_x,k_y)&=\sum_{\mathcal D_N}\tau_{ab}\mathcal A_{ab}(k_x,k_y),\nonumber\\
 Y_N(k_x,k_y)&=X_N(k_y,k_x),\nonumber\\
 m_N(k_x,k_y)&=\mu_0+\sum_{\mathcal D_N}\nu_{ab}\mathcal B_{ab}(k_x,k_y),
 \label{eq:sectors}
\end{align}
where the sums run over $(a,b)\in\mathcal D_N$ and
% APS display rows: 2
\begin{align}
 \mathcal A_{ab}&=\sin(ak_x)\cos(bk_y)+\sin(bk_x)\cos(ak_y),\nonumber\\
 \mathcal B_{ab}&=\cos(ak_x)\cos(bk_y)+\cos(bk_x)\cos(ak_y).
 \label{eq:harmonics}
\end{align}
The equal coefficients within each mixed harmonic define the
isotropic family. Each retained orbit has independent real
hybridization and mass amplitudes, $\tau_{ab}$ and $\nu_{ab}$,
respectively. The two sectors have the same spatial support, and
no common sign or monotonic decay is imposed on their amplitudes.
The first five representatives are $(1,0),(1,1),(2,0),(2,1),(2,2)$.
At shell thirteen, $(5,0)$ and $(4,3)$ enter independently, giving
fourteen orbit amplitudes per sector [Fig.~\ref{fig:support}(a)].
The complete displacement-orbit convention is specified in
Sec.~SI of SM~\cite{SupplementalMaterial}. Section~SI.A relates
the present normalization to that of the supplementary calculations.
\revstop

\revstart
The energies are $E_\pm=\pm\sqrt{|g_N|^2+m_N^2}$, so a gap closes
only where hybridization and mass vanish simultaneously. We maximize
the absolute lower-band Chern number over all gapped choices of
these coefficients, including vanishing hopping amplitudes. A scalar term affects neither eigenstates nor the direct gap and is
omitted. The resulting bound applies
to this specified two-orbital family; changing its symmetry relations
or extending the mass support changes the optimization problem.
\revstop

\revstart
The occupied-state orbital pseudospin points opposite to $(X,-Y,m)$.
Its absolute net wrapping number over the Brillouin zone is $|C|$.
Higher harmonics permit additional windings, with opposite orientations
canceling. The mass selects how the texture passes between opposite
orbital polarizations, so its momentum dependence matters as much as
the hybridization zeros.
\revstop

\begin{samepage}
\textit{Dirac points and their masses.---}
The role of the two hopping sectors becomes transparent near a
simple zero $\mathbf k_i$ of $g_N$. Hybridization vanishes there,
and its linear expansion gives the kinetic terms of a Dirac
Hamiltonian. The hybridization-vortex charge is
$q_i=\sgn\det[\partial(X,Y)/\partial(k_x,k_y)]_{\mathbf k_i}=\pm1$.
At that momentum the band separation is $2|m(\mathbf k_i)|$,
with the mass sign selecting
the orbital character of the occupied state at that momentum.
For $N_0$ simple zeros, the lower-band Chern number is~\cite{Sticlet2012}
% APS display rows: 1
\begin{equation}
 C=\frac12\sum_iq_i\sgn m(\mathbf k_i).
 \label{eq:vortex}
\end{equation}
\end{samepage}
Thus the topological contribution of each gapped Dirac point depends
on both its orientation and its mass. The total vorticity vanishes,
$\sum_iq_i=0$, so a mass with the same sign at every zero gives
$C=0$. A nonzero Hall response requires the mass to distinguish
between vortices of opposite charge. The sign convention and
derivation of Eq.~\eqref{eq:vortex} are given in Sec.~SII of
SM~\cite{SupplementalMaterial}.

\revstart
If the mass could be chosen independently at every zero, setting
$\sgn m_i=q_i$ would give $C=N_0/2$. Finite hopping support links
these choices: all masses are values of the same trigonometric
polynomial in Eq.~\eqref{eq:sectors}. A mismatched sign changes one
contribution from $+1/2$ to $-1/2$ and lowers $C$ by one. Denoting
by $\eta(\mathbf t)$ the minimum number of mismatches over all gapped
mass choices at fixed hybridization coefficients $\mathbf t$, we obtain
% APS display rows: 1
\begin{equation}
 {|C|_{\max}[\mathbf t]=\frac{N_0(\mathbf t)}2-\eta(\mathbf t).}
 \label{eq:fixed}
\end{equation}
Here $|C|_{\max}[\mathbf t]$ maximizes $|C|$ at fixed $\mathbf t$, and
equals the maximum of $C$ because reversing every mass coefficient
reverses $C$. Furthermore, symmetry-related zeros with identical
mass and vorticity can be grouped, with their multiplicity as the
weight of a mismatch (Sec.~SII.A of SM~\cite{SupplementalMaterial}).
\revstop

\revstart
The second-neighbor model gives a simple example that can be checked
directly. When $u=-\tau_{10}/(2\tau_{11})$ satisfies $|u|<1$,
the four zeros at $\Gamma=(0,0)$, $X=(\pi,0)$, $Y=(0,\pi)$,
and $M=(\pi,\pi)$ have $q=+1$. Four additional zeros at
$(\pm k_0,\pm k_0)$, with $\cos k_0=u$, have $q=-1$ and a
common mass $m_{\rm off}$. Equation~\eqref{eq:sectors} gives
% APS display rows: 1
\begin{equation*}
 m_{\rm off}=\frac{(1+u)^2}{4}m_\Gamma
 +\frac{1-u^2}{2}m_X+\frac{(1-u)^2}{4}m_M,
\end{equation*}
where $m_X=m_Y$. The three weights are positive and sum to one:
$m_{\rm off}$ cannot oppose all three masses. Thus the sign pattern
needed for $|C|=4$ is impossible, and $|C|\le3$. Outside this regime there are only four simple zeros; the bound
also holds when zeros merge, since the Chern number remains
unchanged as long as the band gap stays open. The bound is saturated, as shown in Sec.~SIV.C of
SM~\cite{SupplementalMaterial}. Adding Dirac points therefore does not give independent control
of their contributions to the Chern number.
\revstop

\begin{figure*}[t]
\revstart
\centering
\includegraphics[width=\textwidth]{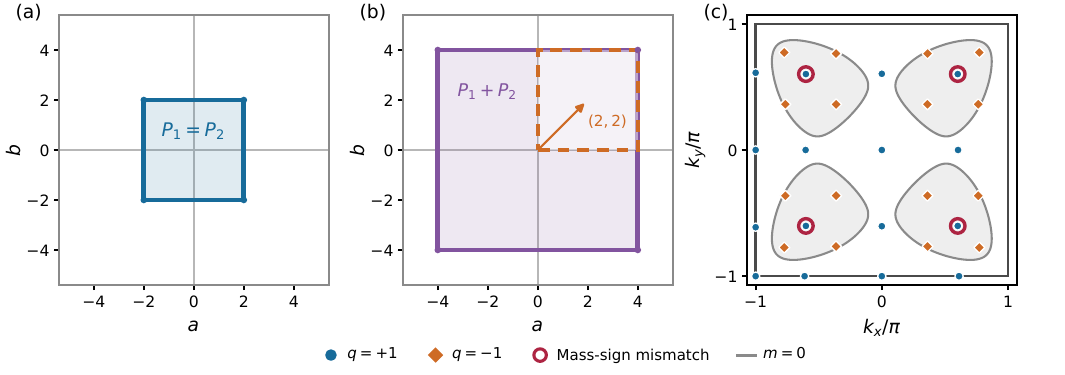}
\caption{From hopping exponents to Dirac points for the
fifth-neighbor Hamiltonian in Eq.~\eqref{eq:n5_witness}.
(a) The hybridization components $f_1=X$ and $f_2=Y$ have
Newton polygons $P_1=P_2=[-2,2]^2$, each of area $16$.
Dots mark vertices; $(a,b)$ labels the harmonic $e^{i(ak_x+bk_y)}$.
(b) Their Minkowski sum is $[-4,4]^2$, of area $64$.
The dashed square shows $P_1+(2,2)$, with the translation indicated
by the arrow. Bernshtein's theorem, Eq.~\eqref{eq:mixed}, gives
the generic complex zero count $D=64-16-16=32$.
(c) All $N_0=32$ zeros are simple and physical, shown once on
$[-\pi,\pi)^2$. Blue circles and orange diamonds denote vortex
charges $q=+1$ and $-1$; gray regions have $m<0$, bounded by $m=0$.
Red rings identify $\eta=4$ unfavorable mass signs, so $|C|=32/2-4=12$, the fifth-neighbor maximum
of Eq.~\eqref{eq:bound}. Roots and mass signs are given in Sec.~SVII.D of
SM~\cite{SupplementalMaterial}.}
\label{fig:mechanism}
\end{figure*}

\revstart
To illustrate how these mass constraints limit the Chern number
at longer hopping range, Fig.~\ref{fig:mechanism}(c) shows the
fifth-neighbor Hamiltonian
% APS display rows: 2
\begin{align}
 \boldsymbol\tau&=(2,8/5,3,26/5,4),\nonumber\\
 (\mu_0;\boldsymbol\nu)&=(1;2,-2,10,4,-2).
 \label{eq:n5_witness}
\end{align}
Its $32$ simple hybridization zeros would give $C=16$ if every
mass sign matched its vortex charge. Here the negative-mass
regions contain all negative-vorticity zeros and four
positive-vorticity zeros (red rings), giving $C=16-4=12$.
This example connects the mass-sign picture to the global bound
derived below, which proves that $12$ is the fifth-neighbor
maximum. Section~SVII.D of SM~\cite{SupplementalMaterial}
gives the roots and mass signs.
\revstop

\textit{Determining the allowed mass signs.---}
At fixed hybridization, changing the onsite imbalance and diagonal
hoppings tunes the Dirac masses without moving the zeros. These
masses share the same amplitudes, so their signs must be chosen
collectively. Write
$m_i=\boldsymbol\Phi_i\cdot\boldsymbol\mu$, where
$\boldsymbol\Phi_i=(1,\boldsymbol\phi_i)$ contains the allowed mass
harmonics evaluated at $\mathbf k_i$, and $\boldsymbol\mu$ contains
the onsite and hopping coefficients. The resulting restriction becomes explicit when a group
$\mathcal C$ of Dirac points satisfies
% APS display rows: 1
\begin{equation}
 \sum_{i\in\mathcal C}\lambda_iq_i\boldsymbol\Phi_i=0,
 \qquad \lambda_i>0.
 \label{eq:circuit}
\end{equation}
For any choice of amplitudes, this relation implies
$\sum_{i\in\mathcal C}\lambda_iq_i m_i=0$.
Thus the masses cannot align with every vortex charge in
$\mathcal C$: at least one point must contribute with the
opposite  sign. This is a direct consequence of the shared
hopping amplitudes (Sec.~SII.A of SM~\cite{SupplementalMaterial}).

Maximizing the Chern number requires the fewest mass signs
opposing their vortex charges. For a proposed mismatch set $S$,
we seek coefficients $\boldsymbol\mu$ with
$q_im_i\ge1$ for $i\notin S$ and $q_im_i\le-1$ for $i\in S$; the unit
margin only fixes the mass scale. At fixed hybridization, testing sets
in increasing number of points finds the minimum mismatch
$\eta$ and coefficients that realize it~\cite{Schrijver1986}
(Sec.~SII.B of SM~\cite{SupplementalMaterial}).
We next establish a bound valid for all gapped choices of
hybridization and mass coefficients, thereby determining the
maximum Chern number at fixed hopping range.

\revstart
\textit{Global Chern-number bound.---}
At larger hopping ranges, the mass constraints must remain valid
as the Dirac points move. Sum rules provide this control:
sufficiently short Fourier harmonics have vanishing weighted sums
over the hybridization zeros. At physical Dirac points, the weight
is proportional to the inverse velocity determinant; complex zeros
also enter and must be retained until $C$ is extracted.
\revstop

\revstart
A displacement $(a,b)$ contributes
$e^{i(ak_x+bk_y)}=z^aw^b$, with $z=e^{ik_x}$ and $w=e^{ik_y}$.
The Fourier exponents of the real hybridization components $f_1,f_2$
are therefore real-space displacement vectors; their enclosing
polygons $P_1,P_2$ describe the spatial extent of these hoppings.
We use $X,Y$ for $N=5,6$ and $X+Y,X-Y$ for $N\ge7$;
the latter combinations cancel harmonics while preserving the
common zeros and $|C|$. Allowing $z,w$ to be nonzero complex
variables, Bernshtein's theorem converts polygon areas into the
generic zero count~\cite{Bernstein1975}
% APS display rows: 1
\begin{equation}
 D=\operatorname{Area}(P_1+P_2)
   -\operatorname{Area}(P_1)-\operatorname{Area}(P_2).
 \label{eq:mixed}
\end{equation}
Fourier multiplication adds exponent vectors.
Figure~\ref{fig:mechanism}(a,b) illustrates this for
$P_1=P_2=[-2,2]^2$: translating one square over the other fills
$[-4,4]^2$, giving $D=64-16-16=32$ in Euclidean areas.
Physical roots satisfy $|z|=|w|=1$, as all $32$ do for the
illustrated Hamiltonian [panel (c)]. Generally, some roots are
nonphysical; $D_{\rm fin}$ counts all finite complex roots.
\revstop

\revstart
Equation~\eqref{eq:vortex} counts the imbalance of positive and
negative mass--vorticity contributions. To constrain this imbalance,
we collect the weights into a Hermitian form. Let $Z_{\rm fin}$ be
the $D_{\rm fin}$ common zeros of $f_1,f_2$ at finite, nonzero
complex $z,w$, including physical zeros on $|z|=|w|=1$.
Initially assume simple zeros and nonzero masses at every root.
Auxiliary finite Fourier sums $\psi,\chi$ assign weights to these
zeros; suitable choices select individual roots. Define~\cite{Pedersen1993}
% APS display rows: 1
\begin{equation}
 \mathcal B(\psi,\chi)=\sum_{\zeta\in Z_{\rm fin}}
       \frac{m(\zeta)\psi(\zeta)\chi^*(\zeta)}{J(\zeta)},
 \label{eq:residue}
\end{equation}
where $J=\det[\partial(f_1,f_2)/\partial(\log z,\log w)]$
and $\chi^*(z,w)=\overline{\chi(1/\bar z,1/\bar w)}$.
In a basis selecting individual roots, each physical zero gives a
diagonal entry $m/J$. On the physical torus,
$J=-\det[\partial(f_1,f_2)/\partial(k_x,k_y)]$, so $m/J$ has
the mass--vorticity sign up to a common orientation.
The star couples nonphysical roots related by
$(z,w)\mapsto(1/\bar z,1/\bar w)$ into off-diagonal blocks with
one eigenvalue of each sign. Let $n_\pm$ count the positive and negative eigenvalues of
$\mathcal B$, with multiplicity. Nonphysical pairs cancel
from their difference, giving $|n_+-n_-|=2|C|$, while
nondegeneracy ensures $n_++n_-=D_{\rm fin}$.
\revstop

\revstart
The remaining task is to establish a guaranteed lower bound $r$
on both $n_+$ and $n_-$. A standard property of nondegenerate
Hermitian forms provides the route: an $r$-dimensional space on
which every pairing vanishes requires at least $r$ eigenvalues
of each sign~\cite{Rodman2008}. Indeed, projection onto either
sign sector must preserve independence; otherwise a nonzero
vector would lie entirely in the other sector and could not
have zero self-pairing. Consequently,
% APS display rows: 1
\begin{equation}
 {|C|\le \frac{D_{\rm fin}}{2}-r.}
 \label{eq:bound}
\end{equation}
We therefore construct independent Fourier probes whose
\emph{every mutual pairing}, including self-pairings, vanishes.
Their independence is required on the zero set, where the form acts.
Section~SXVI.A of SM~\cite{SupplementalMaterial} gives the extended proof.
\revstop

\revstart
Finite hopping range supplies a direct construction because it
restricts the mass harmonics. Choose probes
$\varphi_{\mathbf u}=e^{i(u_xk_x+u_yk_y)}$ with integer
vectors $\mathbf u\in Q$. Their pairings have numerator
$m\varphi_{\mathbf u}\varphi_{\mathbf v}^{*}
=m e^{i(\mathbf u-\mathbf v)\cdot\mathbf k}$:
each mass harmonic is shifted by $\mathbf u-\mathbf v$.
Thus, if $M$ is the mass polygon, the condition
% APS display rows: 1
\begin{equation*}
 M+(Q-Q)\subset\operatorname{int}(P_1+P_2)
\end{equation*}
places every shifted harmonic strictly inside the combined
hybridization polygon. Here
$Q-Q=\{\mathbf u-\mathbf v:\mathbf u,\mathbf v\in Q\}$.
When the finite simple roots exhaust the Bernshtein count,
the toric Euler--Jacobi identity~\cite{Soprunov2006} then gives
$\mathcal B(\varphi_{\mathbf u},\varphi_{\mathbf v})=0$
for every pair, summing over physical and nonphysical roots alike. Separate
rank checks establish $r$ independent evaluated probes.
\revstop

\revstart
The fifth shell makes this connection explicit, see also Fig.~\ref{fig:mechanism}. Here
$P_1=P_2=M=[-2,2]^2$. Choosing
$Q=\{(0,0),(1,0),(0,1),(1,1)\}$ gives the four probes
$1,e^{ik_x},e^{ik_y},e^{i(k_x+k_y)}$.
Since $Q-Q=\{-1,0,1\}^2$, their pairings shift the mass
harmonics into $[-3,3]^2$, strictly inside
$P_1+P_2=[-4,4]^2$. The residue identity therefore makes
every mutual pairing vanish. Evaluation at the four momenta
$k_x,k_y\in\{0,\pi\}$ gives an invertible matrix, establishing
four independent probes and hence $r=4$.
With $D_{\rm fin}=32$, Eq.~\eqref{eq:bound} gives
$|C|\le16-4=12$, saturated by the Hamiltonian in
Fig.~\ref{fig:mechanism}(c).
\revstop

At later shells, additional sum rules appear after removing outer
Fourier terms: write $H=H_{\rm int}+Af_1+Bf_2$ with $H_{\rm int}$
strictly interior. Because $f_1=f_2=0$ at every common zero,
$H$ and $H_{\rm int}$ have the same residue sum, which vanishes.
The support inclusion suffices for $N=5,6,7,12$; a monomial
subtraction at $N=11$ also needs no coefficient relation.
The $N=8,9,13$ reductions prescribe outer-face shapes while leaving
lower coefficients independent. At $N=10$, the interior-support
construction requires an additional boundary calculation in the
original hopping family. The precise coefficient conditions are
given in Secs.~SXVI.C--G of SM~\cite{SupplementalMaterial}.
Here $D=80$, but the prescribed outermost harmonics place four
intersections at the complex boundary, leaving $D_{\rm fin}=76$.
A generic perturbation brings all $80$ roots into the finite
domain, where the residue identity applies. As the perturbation
is removed, the local mass numerator vanishes at these intersections,
making the four escaping residues tend to zero (Sec.~SXVI.E).
The remaining $76$ roots therefore satisfy the same sum rules.
The $r=8$ independent probes give $|C|\le76/2-8=30$;
this improvement uses the coefficient relations of the original family.

Probe independence is established in a simple axial reference
configuration and survives as the Dirac points move under small
changes of the hopping amplitudes; any additional Dirac points only
add constraints and cannot remove this independence. The bound
therefore persists under continuous deformations of the Hamiltonian.
As long as the bulk gap remains open, the Chern number is unchanged,
so one may slightly shift the onsite mass to avoid accidental
band touchings without changing the phase. This continuously connects
the regular configurations used in the proof to cases with vanishing
hoppings or merged Dirac points. Details of this continuation and the
range-by-range checks are given in Secs.~SXVI.B--G of
SM~\cite{SupplementalMaterial}.

\revstart
\textit{Sharp maxima through thirteen shells.---}
Table~\ref{tab:bounds} gives the residue dimensions and maximal
Chern numbers for $N=5,\ldots,13$, each with an explicit
saturating Hamiltonian. Together with the maxima $1,3,6,8$ for the first four
levels of the neighbor hierarchy, these results establish
$|C|_{\max}^{(N)}$ through thirteenth-neighbor hopping.
Proofs and saturating Hamiltonians are given in Secs.~SIII--SVI
and SXVI and Table~S3 of SM~\cite{SupplementalMaterial}.
% APS table rows: 10
\begin{table}
\renewcommand{\arraystretch}{1.3}
\resizebox{.5\columnwidth}{!}{%
\begin{tabular}{|c|c|c|c|c|}
\hline
$N$ & $D_{\rm fin}$ & $r$ & $|C|_{\max}^{(N)}$ & $g_{\rm num}/t_\ast$\\
\hline
5  & 32  & 4  & 12 & $1.4\times10^{-1}$\\
\hline
6  & 48  & 7  & 17 & $1.7\times10^{-2}$\\
\hline
7  & 48  & 5  & 19 & $2.1\times10^{-3}$\\
\hline
8  & 60  & 7  & 23 & $4.3\times10^{-3}$\\
\hline
9  & 80  & 12 & 28 & $4.2\times10^{-2}$\\
\hline
10 & 76  & 8  & 30 & $1.7\times10^{-3}$\\
\hline
11 & 96  & 12 & 36 & $1.2\times10^{-3}$\\
\hline
12 & 96  & 8  & 40 & $9.5\times10^{-3}$\\
\hline
13 & 140 & 19 & 51 & $7.4\times10^{-5}$\\
\hline
\end{tabular}%
}
\caption{Sharp bounds from the hopping support. %
$|C|_{\max}^{(N)}=D_{\rm fin}/2-r$ is saturated in every row.
$D_{\rm fin}$ counts finite complex roots in the proof. The numerical minimum-gap
estimate $g_{\rm num}/t_\ast$ refers to the listed saturating Hamiltonian,
with $t_\ast=\max\{|\tau_{ab}|,|\nu_{ab}|,|\mu_0|\}$.
These diagnostics are independent of the proof.
See Sec.~SXVI, Table~S3, and Sec.~SXVIII of
SM~\cite{SupplementalMaterial}.}
\label{tab:bounds}
\end{table}
\revstop

\revstart
The bounds in Table~\ref{tab:bounds} count all finite complex zeros,
whereas the Chern number receives contributions only from physical
Dirac points. At $N=6$, the bound gives
$|C|_{\max}=48/2-7=17$, while the saturating Hamiltonian realizes
the same value through $44$ physical zeros and five mass-sign
mismatches: $|C|=44/2-5=17$
(Sec.~SVIII of SM~\cite{SupplementalMaterial}).
The remaining four complex zeros form two nonphysical pairs,
each contributing opposite signs to the Hermitian form and hence
zero net signature. For any Hamiltonian saturating
the bound, $r=\eta+(D_{\rm fin}-N_0)/2$; here $7=5+2$, the two units
beyond the mismatches coming from the two nonphysical pairs.

The role of the mass constraints is especially clear at $N=11,12$.
Both saturating Hamiltonians have $88$ physical zeros, but their
allowed mass harmonics give minimum mismatch counts of $8$ and $4$,
respectively, yielding $|C|=36$ and $40$
(Secs.~SXIII--SXIV of SM~\cite{SupplementalMaterial}).
The larger Chern number therefore comes from better alignment
of the mass signs with the vortex charges, despite an equal
number of Dirac points.
\revstop

\revstart
At $N=13$, two inequivalent outer orbits occur at the same distance.
Their independent amplitudes give component and sum polygons of
areas $62$, $62$, and $264$, respectively, hence $D=D_{\rm fin}=140$.
Two successive cancellations of outer Fourier terms give a space
of mutually vanishing pairings: thirteen monomials with
$|a|+|b|\le2$, four individual corner-adjacent monomials, and two
sums whose remaining pairings cancel. Their independent evaluations
give $r=13+4+2=19$, hence $|C|\le51$.
See Sec.~SXVI.G of SM~\cite{SupplementalMaterial} for the support geometry and probe construction.
A Hamiltonian with $116$ simple physical zeros and seven
mismatches saturates the bound at $C=58-7=51$ (Sec.~SXV.B). Every maximum through $N=12$ satisfies
$|C|_{\max}^{(N)}\le2\rho_N$, with equality at $N=11$ and $12$
[Fig.~\ref{fig:support}(b)]. The thirteenth shell, the first with two
inequivalent orbits, breaks this envelope:
$|C|_{\max}^{(13)}=51>2\rho_{13}=50$.
For any generic configuration of zeros at this shell,
Eq.~\eqref{eq:fixed} also yields the uniform constraint
$\eta\ge N_0/2-51$ (Sec.~SXV.C).
\revstop

\revstart
For each saturating Hamiltonian, root isolation together with
rigorous bounds on the Dirac-point locations, vorticities, and masses
certifies the Chern number and a finite bulk gap, while a discretized
Brillouin-zone calculation~\cite{Fukui2005} provides an independent
numerical check (Secs.~SXVI.H and SXVII of
SM~\cite{SupplementalMaterial}).

\textit{Discussion and outlook.---}
Finite hopping range constrains both the number of Dirac points
and the mass patterns that gap them. Because these masses arise
from shared hopping amplitudes, additional zeros do not provide
independent  contributions to the Chern number. The exact sum rules quantify this
restriction and establish global bounds that are saturated through
thirteenth-neighbor hopping in the specified square-lattice family.

This perspective is particularly relevant to moir\'e systems, where
extended orbitals and interference can make further-neighbor hopping
compete with shorter-range processes~
\cite{Koshino2018,Crepel2024,Eugenio2025}. Higher-Chern moir\'e
flat bands provide a natural setting in which to apply these
constraints~\cite{Ledwith2022,WangHighC2025}.
The appropriate lattice, hopping support, and microscopic coefficient
relations determine which mass patterns are accessible. Applying the
present framework to such effective models would therefore connect
their hopping hierarchy to the largest attainable Chern number.

Higher-Chern moir\'e bands are especially interesting in the interacting
regime, where they can support fractional Chern insulators, quantum Hall
ferromagnets, and translation-breaking topological phases distinct from
their $C=1$ counterparts~
\cite{Dong2023,WangKlevtsovLiu2023,PereaCausin2025,
WangHighC2025,LiHighC2026}. This further motivates identifying the
microscopic hopping constraints that control how large a Chern number
can be realized.

Experimental relevance also depends on the size of the bulk gap.
For the saturating Hamiltonians, the gaps measured in units of the
largest Hamiltonian coefficient $t_\ast$ vary strongly and
nonmonotonically, from about $10^{-1}$ to below $10^{-4}$
(Table~\ref{tab:bounds}; Sec.~SXVIII of SM~\cite{SupplementalMaterial}).
The saturating phases therefore span a broad range of bulk gaps, while
maximizing the gap within each Chern sector is a separate optimization
problem.

The root-counting and signature framework also extends naturally to
other two-orbital lattices, including honeycomb models~\cite{Haldane1988,Sticlet2013},
where different hopping geometries and coefficient relations lead to
new sum rules and saturating phases. Extending the approach to
generic multiband systems will require a replacement for the
single-pseudospin structure underlying the present signature relation.
An exact expression for the maximal Chern number at arbitrary hopping
range also remains open. Such a formula should encode the key physical
constraint identified here: how effectively finite-range hopping allows
the mass sector to distinguish vortices of opposite charge.

\paragraph*{Acknowledgments.} This work is supported by Fondecyt (Chile) Grant  No. 1230933 (V.J.) and PIIC scholarship (Programa de Iniciación a la Investigación Científica) under Agreement No. 023/2025 (C.A.C.). We thank Bitan Roy for useful discussions. 

\paragraph*{Declaration of AI use.}
ChatGPT (OpenAI) was used to assist with the manuscript editing  and calculations. 

%\paragraph*{Data availability.}
%All supporting data and certificates accompany the SM source package; see Sec.~SXVII of SM~\cite{SupplementalMaterial}. 

\bibliography{refs}
\end{document}